# A Protection Approach for Video Information transmitted in TCP/IP based networks

Nikolai Stoianov
Technical University Sofia, Bulgaria
nkl_stnv@tu-sofia.bg

*Abstract* — In this paper an analysis of the existing video information protection methods is made. Analysis of current H.323 protocol stack has been made. A new encryption/decryption layer has been suggested. An approach for partial data encryption in the H.323 protocol stack is proposed and sample architecture of a Virtual Private Video Network is given.

## I. INTRODUCTION

With the development of the Information Technologies users are provided with more and more new services. Video conferences with any point of the world for example are already a daily practice. Sending and receiving any kind of information (text, video, voice) is of extreme importance for every organization. One of the most successful and secure protection methods for data transmitted in the computer systems is the use of encryption techniques [8]. The deployment of video data transmission networks is widely used surveillance pattern by the police authorities of European countries. It is necessary those networks and the exchanged data to remain hidden for external and unauthorized users. One of the main problems to be solved by experts is to find the data encryption method and to preserve the functionality of the audio-video protocols.

One of the main goals of each security system is to protect the data from unauthorized access. To achieve this goal in the video information protection system is rather difficult because of the following circumstances:
- The necessity to preserve the functionality of the end users (hardware and software);
- Possibility to exchange data considering the necessary quality level (Quality of Service);
- Adaptability of the human eye and object recognition capability.

## II. CLASSIFICATION OF THE VIDEO INFORMATION ENCRYPTION METHODS

In the literature there are many video information encryption methods and approaches. Some of them use the complete encryption possibility; others encrypt partially only some data based on predefined approaches and criteria [1][2][6].

*Complete Encryption* – This data protection method encrypts the whole multimedia content without a difference of the data format in it.

*Partial Encryption* – These are methods and algorithms that firstly divide all data into two parts and only after encrypt just one part of the content, but the other part remains unprotected.

*Compression-Combined Encryption* – In this method algorithms which encrypt and compress the data are used and as a result of this operation protected data of a smaller size are received. In difference of the traditional scheme of compression and next encryption, in this method the data compression and encryption are done simultaneously.

*Perceptual Encryption* – In this method the protection is achieved via reduction of the quality of the video information, related to the required levels of protection and the quality of the transmitted information.

*Scalable Encryption* – This data encryption method combines the scalable compression with encryption algorithms. As a result of these actions protected data with a lower quality level than the original are generated. In this case the quality requirements can not be specified but it is defined by the compression mode.

*Video Scrambling* – In this method the so-called scrambling of the video signal is used, i.e. mixing the useful signal with the predefined binary sequence. Most often XOR operations are used (sum by module 2) or substitution of the input signal with predefined replacement matrix. This is one relatively fast signal processing method, but at the same time it is not too reliable. This method is mostly used by the cable TV operators to protect their paid channels.

On Fig. 1 the classification of methods for video information encryption is shown.

## III. ARCHITECTURE OF VIRTUAL PRIVATE VIDEO NETWORK

One of the most spread protocol stacks for audio and video information is H.323. As per the stack specifications a separate protocols for the audio data are defined in it (G.711, G723 and others), and for video data (H.261, H.263), the calls control, signalization, registration, administration and status (RAS) [3][4][5]. In the H.323 stack there is no specific special protection tool (encryption) for the transmitted data. Investigating the inter link between the OSI model and the H.323 it is obvious that it is possible to define an interim layer which encrypts the data (audio and video) and at the same time to preserve the functionality of the protocol. In this way the additional data which are not sensitive for the information protection remain "understandable" for the other protocols and applications. Placing the encryption/decryption layer between the H.225 and audio and video encoding systems provides an opportunity to execute calls, to add flags for the required quality or priority of processing of the signals but only the main data are(is) protected. On Fig. 2 the content of the protocol stack H.323 and the additional information protection layer are shown.

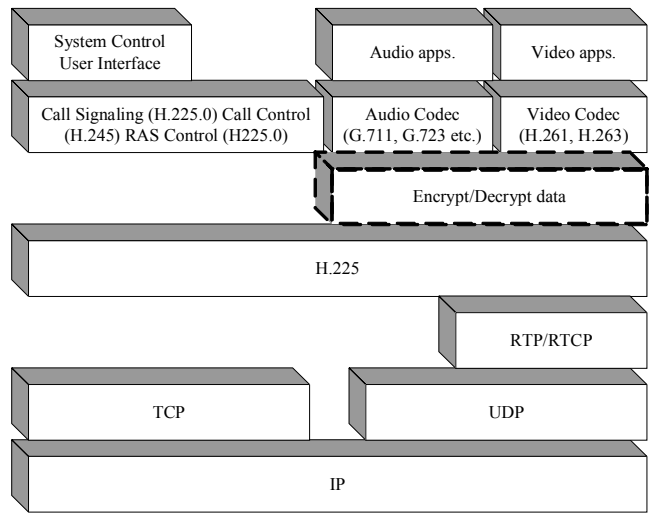

Figure 2. H.323 protocol stack with added Encryption/Decryption layer

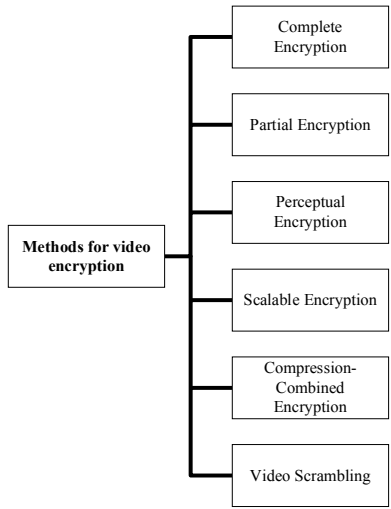

Figure 1. Methods for video encryption

Using the standard way for creating Virtual Private Network in order to protect the audio and video information in tunnel regime, the added to H.225, H.245 and other protocol service information related to signalization and data management, remain hidden for the medium retransmission equipment (routers, switches). Using this data protection pattern via the classic tunnel VPN (point to point or gateway-gateway), the larger part of the functionality of the signaling protocols is lost. When the data are transmitted in real time and a routing of the multimedia data is a priority, then this approach in inapplicable via VPN.

Having in mind the necessity to create such type of communication and data protection the so-called Virtual Private Video Network (VPVN) are proposed, where based on the additional encrypting/decrypting layer in H.323 (fig.2) a possibility to create an audio and video information transmission protected network is provided.

The possible solutions for such VPVN are:
- Point-to-point type network;
- Site-to-site type network (crypto gateway-to-crypto gateway);
- Point-to-site type network (crypto gateway).

*Point-to-point type network*

In this type of virtual network the end user devices (hardware and software) conduct the initialization of the session, agreement on the necessary data encryption/decryption keys and the instant protection of the multimedia information. This type of VPVN organization is possible in case of relatively small number of end users or in case of deployment of networks for some specific needs (police, civil protection) for temporary use (fig. 3). A similar type of network can be also created for direct communication between mobile users for example police cars or police car and ambulance or fire brigade vehicle.

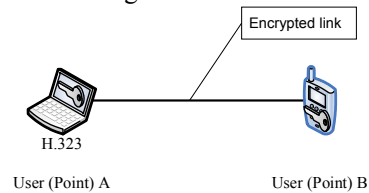

Figure 3. Point-to-point VPVN

*Site-to-site type network (crypto gateway-to-crypto gateway)*

Architecture of the site-to-site type allows multiple users to establish (transparent for them) audio and video links with users of other site. In this way of creating an encrypted link functionally new cryptographic devices are added (crypto gateway), which build among them an encrypted video channel (fig. 4). It is reasonable such type of architecture to be used for establishing video link between many subscribers, located in one and the same building (police station - department).

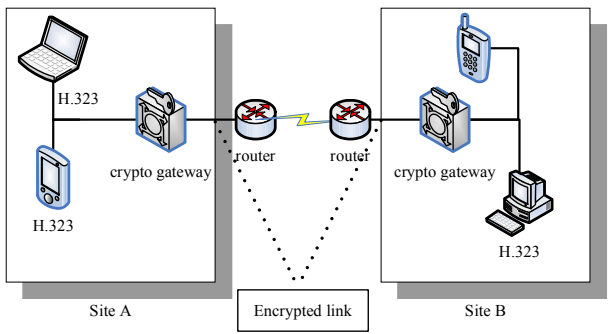

Figure 4.  Site-to-site VPVN

*Point-to-site type network*

In this type of network the possibility for direct communication between one distant (mobile) user and centralized access point. In this case the initiator of the connection is normally the user of point type (fig. 5). The user sends to the site a request for multimedia connection. If the user has access rights to the crypto gateway, he conducts an encrypting of audio and video information based on the additional functionality via encrypting/decrypting level in H.323 protocol stack. This architecture is possible to be used in the link between a police car and coordination center, ambulance and emergency departments.

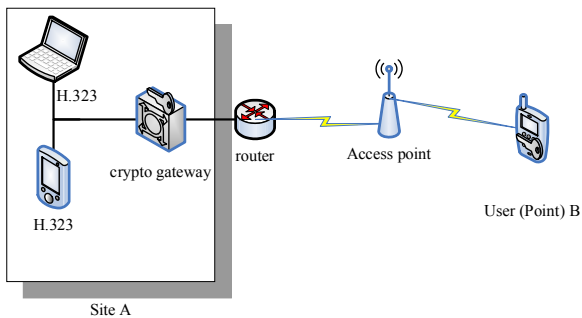

Figure 5.  Point-to-site VPVN

All the proposed VPVN architecture types provide data protection/encryption possibilities and at the same time to add flags for processing priorities (routing) of packages, signalization control and determination of different levels of the quality of services (QoS).

In the proposed VPVN architectures it is possible to use block and stream encryption algorithms. The encryption method is so-called "complete encryption", i.e. all the multimedia data are protected. It is recommended the AES-256 to be used.

The main question in the deployment of cryptographic protection system is the way of negotiating (exchange) of cryptographic keys. In the proposed VPVN architecture it is reasonable to use the combined method for negotiation of keys, namely the elliptic curve cryptography (ECC) is used to negotiate the session keys for the AES, via which the multimedia data are protected. In order to unify the architecture, an independent module for cryptographic session keys – Key Distribution Centre (KDC) is added. The role of this module is in case of request for encrypted video session between two subscribers (no matter if they are point type or crypto gateway) using a secure channel between the subscriber and the KDC, the request is encrypted with an asymmetric algorithm based on ECC. KDC receives the request for new key generation, generates temporary session key and sends it to the source and destination address. After receiving the temporary session key, the source and destination use it for encrypting/decrypting of the multimedia data based on the modified protocol stack H.323 (fig. 2).

IV.  E-NETWORK MODEL OF VPVN WITH KDC

For modeling the interaction between the VPVN users (of point and crypto gateway types) a network with 5 subscribers is defined (three crypto gateway and two point type) (fig. 6). The network provides the data protection via AES-256 algorithm. Each subscriber has public and private key (the asymmetric algorithm is based on ECC). The negotiation of the protection keys for each session are used temporary (session) keys generated by the Key Distribution Centre (KDC).

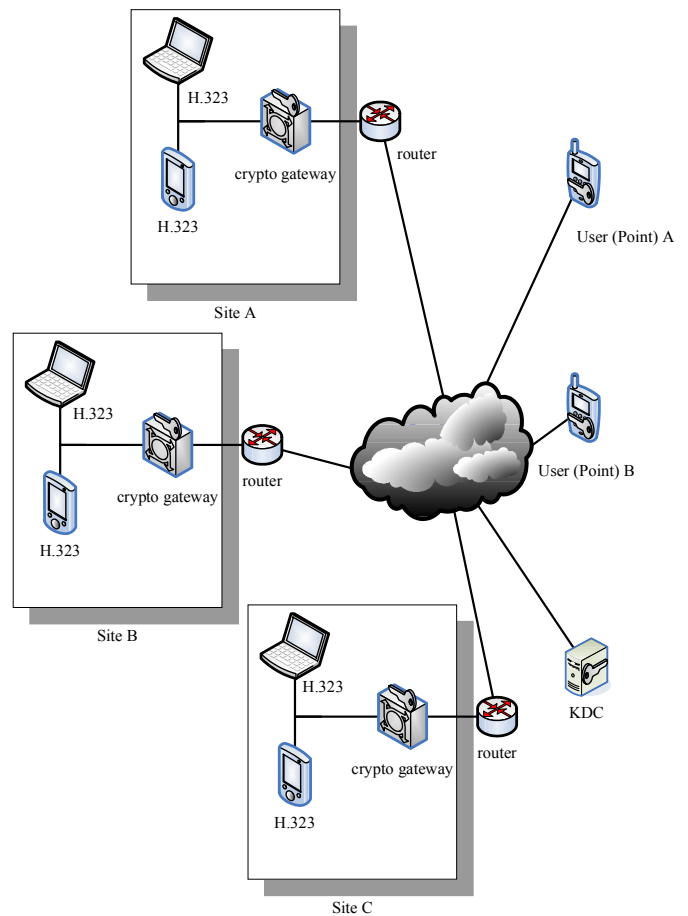

Figure 6.  Sample VPVN with KDC

The functional possibilities for establishing encrypted multimedia connection are:
- Point A (Point B) to Site A (Site B or Site C);

- Site A to Site B (Site C);
- Site A to Site B to Site C;
- Site A (Site B, Site C) to Point A (Point B);
- Point A to Point B.

In each of those combinations the KDC is used (fig. 6) for distribution and negotiation of the data protection cryptographic keys.

For such defined VPVN, using the apparatus of the E-networks [7][8], a E-network model EN_VPVN=<B, Bp, Br, T, F, H, Mo> is created, and thru it the interaction between all the users will be evaluated, where (fig. 7):

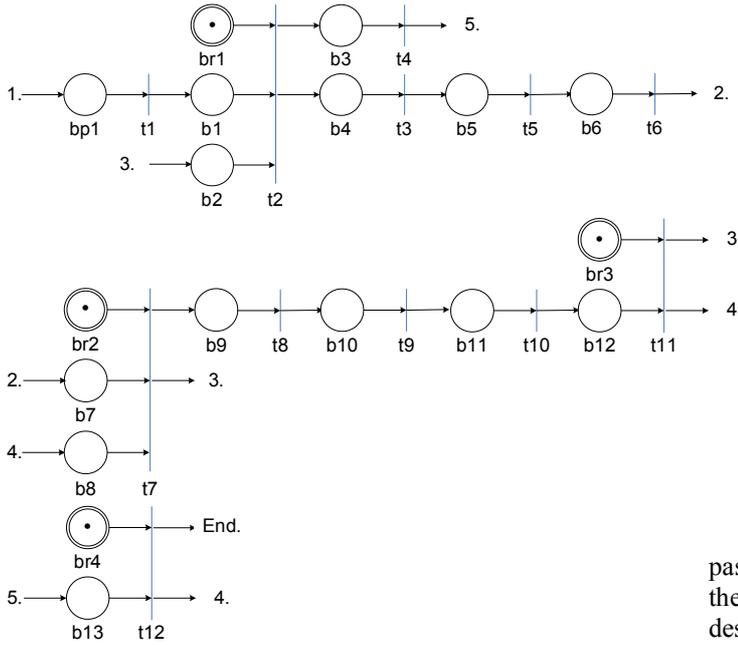

Figure 7. E-net model EN_VPVN for VPVN with KDC

The positions of the model describe the status and the interaction of the users with the system.

B= {bp1, br1,..., br4, b1,..., b11} is the multitude of the positions of the model;

Bp={bp1} is the multitude of the peripheral positions, where in bp1 a core appears exactly when an user (point or crypto gateway) sends a request for initiation of new secure session to the KDC.

Br= {br1, br2, br3, br4} is the multitude of allowing positions respectively for the transitions t2, t7, t11 and t12, where:

br1 – defines the logical transition during the authorization check of the initiating side in the KDC;

br2 – defines the logical transition during the check of the necessity to change the session key before the encryption operation;

br3 – defines the logical transition during the check of the necessity to change the session key after the decryption operation;

br4 – defines the logical transition during the check for the end of work;

$F: B \times T \to \{0,1\}$;
$H: T \times B \to \{0,1\}$;
Mo: $B \to \{0,1\}$ is a marking of the positions.
T= {t1,..., t12} is the multitude of the transitions.

The passages of the EN_VPVN model the functioning of the system during the distribution, access and use of cryptographic resources as follows:

The transition t1 models a request for access to the system, the primitive Ident;

t2 – is check of the authorization, primitive CheckAuthorities;

t3 – is a generation of new session key, primitive GenSKey;

t4 – is the rejection of the request for initialization of a session, primitive EndInit;

t5 – is a protection of the generated session key, primitive ECCSKey;

t6 – is sending of the protected key to the source and destination address, primitive SendSKey;

t7 – is data encrypting and check of the necessity for a new session key, Encrypt;

t8 – is sending data to the recipient, Send;

t9 – is receiving data by the recipient, Receive;

t10 – is the decryption of the received data, Decrypt;

t11 – is check of the necessity for a new session key, CheckSKey;

t12 – is exit of the system, primitive Quit.

The functioning of the model is demonstrated in the passage thru sequence of transitions depending on the status of the allowing positions br1,...,br4 and the parameters of the described cores, namely:

t1:[1;0|0,1];
t2:t2` or t2``, where:
    t2`:[0*;0(1);0(1);0;0|0*;0,0;0;1]
    t2``:[1*;0(1);0(1);0;0|0*0;0;1;0]
t3:[1;0|0;1];
t4:[1;0|0;1];
t5:[1;0|0;1];
t6:[1;0|0;1];
t7:t7` or t7``, where:
    t7`:[0*;0(1);0(1);0;0|0*;0,0;1;0];
    t7``:[1*;0(1);0(1);0;0|0*;0;0;0;1];
t8:[1;0|0;1];
t9:[1;0|0;1];
t10:[1;0|0;1];
t11:t11` or t11``, where:
    t11`:[0*;1;0;0|0*;0;1;0];
    t11``:[1*;1;0;0|0*;0;0;1];
t12:t12` or t12``, where:
    t12`:[0*;1;end;0|0*;0;end;1];
    t12``:[1*;1;end;0|0*;0;end*;0].

Sequentially EN_VPVN=t1 ∩ (t2` U t2``) ∩ t3 ∩ t4 ∩ t5 ∩ t6 ∩ (t7` U t7``) ∩ t8 ∩ t9 ∩ t10 ∩ (t11` U t11``) ∩ (t12` U t12``).

## V. CONCLUSION

In this paper one approach for development of protected environment for multimedia information exchange. Based on the protocol stack H.323 it is proposed to add a new layer for data encryption/decryption. Hierarchically this layer is between the audio and video information processing applications and the encoding systems. Via this approach the possibility to receive and process the initialization signals, to control, prioritize the packages and to introduce QoS, as well as to execute a complete encryption of the data at the same time. The proposed architecture for a Virtual Private Video Network gives the possibility to develop a common environment for protected audio and video information transmission. The created e-network model for architecture research is constructive and invariant and assists for the description and analysis of the interaction between the users (point or crypto gateway). The author believes that the described approach is possible for realization not only in the INDECT Project, but also in other multimedia information protection related projects.

## AUTHOR PROFILE

Dr. Nikolai T. Stoianov, more than 10 years of Research Experience Working. Area of Specialization: Information Systems Security, Security Interactions and Encryption Techniques. E-mail id: nkl_stnv@tu-sofia.bg. He is member of Bulgarian Union of Scientists. He has 45 publications at national and international level journals and conferences.